\documentclass[10pt]{amsart}

\usepackage[utf8]{inputenc}
\usepackage{amsfonts}
\usepackage{amssymb}
\usepackage{amsmath,amsthm}
\usepackage{amscd}
\usepackage{graphics}
\usepackage{cancel}
\usepackage{pdfsync}
\usepackage[british]{babel} 
\usepackage{mathrsfs}
\usepackage{enumitem}
\usepackage{stmaryrd}
\usepackage{mathrsfs}
\usepackage{wasysym}
\usepackage{latexsym}
\usepackage{mathtools}
\usepackage{afterpage}
\usepackage{threeparttable}

\usepackage{empheq,etoolbox}
\usepackage{xcolor}
\usepackage[colorlinks,linkcolor=blue,citecolor=blue,urlcolor=blue]{hyperref}

\numberwithin{equation}{section}

\newcommand{\beq}{\begin{equation}}
\newcommand{\eeq}{\end{equation}}
\newcommand{\bea}{\begin{eqnarray}}
\newcommand{\eea}{\end{eqnarray}}
\newcommand{\nn}{\nonumber}
\newcommand\noi{\noindent}
\newcommand{\tbf}{\textbf}

\newcommand{\rd}{\mathrm{d}}

\newcommand{\vphi}{\varphi}

\newcommand{\bk}{\begin{cases}}
\newcommand{\ek}{\end{cases}}
\newcommand{\bepm}{\begin{pmatrix} }
\newcommand{\epm}{\end{pmatrix}}
\newcommand{\bs}{\boldsymbol}

\newtheorem{definition}{Definition}
\newtheorem{proposition}{Proposition}
\newtheorem{theorem}{Theorem}

\newtheorem{lemma}{Lemma}
\theoremstyle{definition}

\newtheorem{remark}{\textbf{Remark}}

\newtheorem{example}{\textbf{Example}}

\usepackage{enumerate}
\usepackage{float}
\usepackage{cancel}
\usepackage{hyperref}
\usepackage{mathtools}
\usepackage{afterpage}

\allowdisplaybreaks

\begin{document}

\title[$\omega \mathscr{H}$ manifolds for integrable systems in a magnetic field]{Hamiltonian integrable systems in a magnetic field and Symplectic-Haantjes geometry}

\author{Ondřej Kubů}
\address{Czech Technical University in Prague, Faculty of Nuclear Sciences and Physical Engineering, Department of Physics, Břehová 7, 115 19 Prague 1, Czech Republic}
\email{ondrej.kubu@fjfi.cvut.cz}
\author{Daniel Reyes}
\address{Departamento de F\'{\i}sica Te\'{o}rica, Facultad de Ciencias F\'{\i}sicas, Universidad
Complutense de Madrid, 28040 -- Madrid, Spain \\ and Instituto de Ciencias Matem\'aticas, C/ Nicol\'as Cabrera, No 13--15, 28049 Madrid, Spain}
\email{danreyes@ucm.es}
\author{Piergiulio Tempesta}
\address{Departamento de F\'{\i}sica Te\'{o}rica, Facultad de Ciencias F\'{\i}sicas, Universidad
Complutense de Madrid, 28040 -- Madrid, Spain \\  and Instituto de Ciencias Matem\'aticas, C/ Nicol\'as Cabrera, No 13--15, 28049 Madrid, Spain}
\email{ptempest@ucm.es}
\author{Giorgio Tondo}
\address{Dipartimento di Matematica, Informatica e Geoscienze, Universit\`a  degli Studi di Trieste,
piaz.le Europa 1, I--34127 Trieste, Italy (retired)}
\email{tondo@units.it}

\date{Jun 3, 2024}




\begin{abstract}
We investigate the geometry of classical Hamiltonian systems immersed in a magnetic field in three-dimensional Riemannian  configuration spaces. We prove that these systems admit non-trivial symplectic-Haantjes manifolds, which are symplectic manifolds endowed with an algebra of Haantjes (1,1)-tensors. These geometric structures allow us to determine separation variables for known systems algorithmically; besides, the underlying Stäckel geometry is used to construct new families of integrable Hamiltonian models immersed in a magnetic field. 
\end{abstract}

\maketitle

\tableofcontents
\section{Introduction}

The purpose of this article is to study classical integrable and superintegrable Hamiltonian models in a magnetic field employing a new geometric perspective based on the theoretical framework of Haantjes geometry. Haantjes tensors \cite{Haa1955}, which have been introduced as a generalization of Nijenhuis tensors \cite{Nij1951}, can be used to define a natural geometric framework for classical Hamiltonian mechanics. On the one hand, the relevance of our approach resides in the fact that  the geometric structures we propose shed new light on the properties of known physical systems; on the other hand, it allows to define new magnetic models from the underlying geometric structure.

The general class of models we are interested in is described by a Hamiltonian of the form (in units $e=-1,m=1$)
\beq \label{eq:1.1}
H=\frac{1}{2}g^{ij}(p_i+A_i)(p_j+A_j)+V(\bs{q})= \frac{1}{2}g^{ij}p_ip_j +A^ip_i+ U(\bs{q}),
\eeq
which is defined on a Riemannian manifold $(Q,g)$ where $g$ is a Riemannian metric on an $n$-dimensional differentiable manifold $Q$,  and $\bs{q}=(q^{1},\ldots,q^{n})$,  $\bs{p}=(p_{1},\ldots,p_{n})$ are the local coordinates on $T^{*}Q$, which are canonically conjugated. In this article, we shall focus on the case $n=3$. Also, $A$ is a smooth vector field (the vector potential), $B=\nabla \times A$ is the magnetic field associated with the model, $V(\bs{q})$ is the scalar potential and 
\beq \label{eq:1.2}
U:= \frac{1}{2}A^i A_i+V.
\eeq
A finite-dimensional Hamiltonian system is superintegrable if it is Liouville integrable and allows more independent integrals of motion than degrees of freedom \cite{MF1978,TWH04,MPW2013,KKM2018Book}.
The investigation of integrable Hamiltonian models immersed in magnetic fields, or with velocity-dependent potentials, started in the 1960s with the study of certain systems with a magnetic monopole \cite{Zwanziger1968,McIntosh1970,Jackiw1980,Labelle1991,MPW2013}. A more systematic search in the two-dimensional (2D) Euclidean space was initiated in the 1980s with the study of systems admitting first-degree integrals in the momenta \cite{DGRWJMP1985}; for second-degree integrals the analysis was pursued in the 2000s, leading to a partial classification  \cite{McWJMP2000,BWJMP2004,Charest2007,Pucacco2004,PRJMP2005}. The analysis in the three-dimensional (3D) Euclidean space, started with the work \cite{Marchesiello2015}, is still in progress (see e.g. \cite{Hoque2023, KMSJPA2024} and references therein). Separability on curved Riemannian spaces has also been considered (see e.g. \cite{BCRJMP2001}). Recently, in \cite{BNJPA2023} a novel approach for the study of the orbits of superintegrable systems in a magnetic field, based on Lie group theory, has been proposed.

The theory of classical integrable systems has been formulated in a tensorial perspective in \cite{MMQUAD1984} and in \cite{FPMPAG2003} using the Nijenhuis geometry (see also \cite{MMT1988}). In this context, the relevant notion of $\omega N$ manifolds, namely symplectic manifolds endowed with a distinguished Nijenhuis tensor, has been introduced and the theory of separation of variables (SoV)  formulated. More recently, in \cite{MGall13}, the relevance of Haantjes tensors in the theory of integrable systems has been enlightened, and the concept of Haantjes manifold has been introduced. Motivated by this research, in \cite{TT2016SIGMA,TT2022AMPA} the notion of symplectic-Haantjes manifold has been proposed as another natural geometric scenario where the theory of separation of variables for Hamiltonian systems can also be formulated. Essentially, a symplectic-Haantjes (or $\omega \mathscr{H}$)  manifold is a symplectic manifold endowed with a Haantjes algebra of tensor fields, which are required to be compatible with the symplectic form. A comparison between the Haantjes manifold and the $\omega \mathscr{H}$ approaches has been proposed in \cite{KS2019}.
The theory of Haantjes tensors has also been related to the well-known class of hydrodynamic-type systems \cite{FKPRA2006} and with the kinetic gas theory \cite{FPJNS2022}. However, to our knowledge, Hamiltonian models immersed in a magnetic field were not studied before in the context of Nijenhuis or Haantjes geometries. 

The main idea of this work is to study the interplay between magnetic systems and their underlying geometry.
In the first part of the paper, we shall consider families of integrable Hamiltonian systems immersed in a magnetic field, which are of physical interest, and we unravel their underlying geometric structure.
More precisely, we shall construct the symplectic-Haantjes manifolds associated with the magnetic systems under investigation. Thus, we prove that the Darboux-Haantjes (DH) coordinates for these geometric structures provide coordinate charts where the Hamilton-Jacobi (HJ) equations associated with our systems separate into systems of ODEs (\textit{total separability}), or ODEs and PDEs (\textit{partial separability}). Also, we shall prove that there is a direct relation between the Haantjes geometry and the Stäckel theory of separability. 
We stress that, in contrast to the usual methods based on the configuration space separation, the full generality of the Haantjes geometry, directly defined on the phase space, bypasses the necessity of fixing from the very beginning a specific gauge for the associated vector potentials, usually done in an ad hoc manner. Indeed, it is the geometry of the full phase space that naturally determines the gauge needed for separation.

In the second part of the paper, we go back from geometry to magnetic systems. We will suitably generalize the Haantjes and Stäckel geometries determined in the first part in order to construct new families of integrable Hamiltonian systems, defined on Riemannian manifolds, immersed in suitable magnetic fields. 

In summary, we have two main results: first, we prove that the Haantjes geometry is a very useful tool for studying the separability of magnetic systems; second, we use the obtained Haantjes-Stäckel geometric structures to define new large families of integrable models parametrized by arbitrary functions, which after a suitable canonical transformation are interpreted as magnetic systems on curved spaces.

A novel observation is that we can regard our new magnetic models as generalized Stäckel systems in the sense of Arnold-Kozlov-Neistadt \cite{AKN1997}. This important class of Hamiltonian integrable models possesses an interesting geometry \cite{TT2016SIGMA}, which will be fundamental in the construction of the new models.

The article is organized as follows. In Section \ref{sec:2},  we revise some basic notions of the Nijenhuis and Haantjes geometries, in particular the construction of symplectic-Haantjes manifolds. In Section \ref{sec:3}, we derive general systems that are separable in cylindrical coordinates by means of a geometric approach, which also provides us with the associated Haantjes algebras. In section \ref{sec:4}, we analyze thoroughly the case of a system in a constant magnetic field. From its superintegrable structure, we were able to determine three different symplectic-Haantjes manifolds, each related to a different separation coordinate system. Another superintegrable model of physical interest, the helical undulator, is also analyzed. In Section \ref{sec:5}, utilizing the Jacobi-Sklyanin separation equations (SE), we construct the Stäckel geometry associated with these superintegrable models. In Section \ref{sec:6}, we generalize it to define new classes of Hamiltonian models in a magnetic field by solving an inverse problem. 

\section{Haantjes geometry and integrable systems} \label{sec:2}

The purpose of this section is to review some of the basic definitions and results of the theory of symplectic-Haantjes manifolds for Hamiltonian integrable systems, which are crucial for the analysis of magnetic systems. Also, we shall revise the connection between Stäckel and Haantjes geometries, which is instrumental for the construction of new magnetic models presented in Section \ref{sec:6}. A more detailed discussion can be found in the original papers  \cite{Nij1951,Haa1955} and in \cite{TT2021JGP,TT2022AMPA,TT2016SIGMA,RTT2022CNS}.
\subsection{Nijenhuis and Haantjes operators}	
	Let $M$ be a differentiable manifold of dimension $n$, $\mathfrak{X}(M)$ the Lie algebra of vector fields on $M$ and $\boldsymbol{K}:\mathfrak{X}(M)\rightarrow \mathfrak{X}(M)$ a smooth $(1,1)$-tensor field. In the following, all tensors will be considered to be smooth.
\begin{definition} \label{def:N}
The \textit{Nijenhuis torsion} of $\boldsymbol{K}$ is the  vector-valued $2$-form defined by
\begin{equation} \label{eq:Ntorsion}
\tau_ {\boldsymbol{K}} (X,Y):=\boldsymbol{K}^2[X,Y] +[\boldsymbol{K}X,\boldsymbol{K}Y]-\boldsymbol{K}\Big([X,\boldsymbol{K}Y]+[\boldsymbol{K}X,Y]\Big),
\end{equation}
where $X,Y \in \mathfrak{X}(M)$ and $[ \, \cdot \, , \cdot \, ]$ denotes the commutator of two vector fields.
\end{definition}

\begin{definition} \label{def:H}
\noi The \textit{Haantjes torsion} associated with $\boldsymbol{K}$ is the vector-valued $2$-form defined by
\begin{equation} \label{eq:Haan}
\mathcal{H}_{\boldsymbol{K}}(X,Y):=\boldsymbol{K}^2 \tau_{\boldsymbol{K}}(X,Y)+\tau_{\boldsymbol{K}}(\boldsymbol{K}X,\boldsymbol{K}Y)-\boldsymbol{K}\Big(\tau_{\boldsymbol{K}}(X,\boldsymbol{K}Y)+\tau_{\boldsymbol{K}}(\boldsymbol{K}X,Y)\Big).
\end{equation}
\end{definition}

\subsection{Haantjes algebras}
Another crucial notion is that of Haantjes algebras, which has been introduced in \cite{TT2021JGP} and further studied and generalized in \cite{RTT2023JNS}. 
\begin{definition}\label{def:HM}
A Haantjes algebra is a pair    $(M, \mathscr{H})$ satisfying the following properties:
\begin{itemize}
\item
$M$ is a differentiable manifold of dimension $n$;
\item
$ \mathscr{H}$ is a set of Haantjes  operators $\boldsymbol{K}:\mathfrak{X}(M)\to \mathfrak{X}(M)$. Also, they  generate:
\begin{itemize}
\item
a free module over the ring of smooth functions on $M$:
\begin{equation}
\label{eq:Hmod}
\mathcal{H}_{ ( f\boldsymbol{K}_{1} +
                             g\boldsymbol{K}_2 )}(X,Y)= \mathbf{0}
 , \qquad\forall\, X, Y \in \mathfrak{X}(M) , \quad \forall\, f,g \in C^\infty(M)\  ,\quad \forall ~\boldsymbol{K}_1,\boldsymbol{K}_2 \in  \mathscr{H};
\end{equation}
  \item
a ring  w.r.t. the composition operation
\begin{equation}
 \label{eq:Hring}
\mathcal{H}_{( \boldsymbol{K}_1 \, \boldsymbol{K}_2 )}(X,Y)=\mathbf{0} , \qquad
\forall\, \boldsymbol{K}_1,\boldsymbol{K}_2\in  \mathscr{H} , \quad\forall\, X, Y \in \mathfrak{X}(M).
\end{equation}
\end{itemize}
\end{itemize}
If
\begin{equation}
\boldsymbol{K}_1\,\boldsymbol{K}_2=\boldsymbol{K}_2\,\boldsymbol{K}_1 , \quad\qquad\ \forall~\boldsymbol{K}_1,\boldsymbol{K}_2 \in  \mathscr{H},
\end{equation}
the  algebra $(M, \mathscr{H})$ is said to be an Abelian  Haantjes algebra. 
\end{definition}
In essence, we can think of $\mathscr{H}$ as an associative algebra of Haantjes operators.
\begin{remark}
A fundamental fact concerning Abelian Haantjes algebras is the existence of a web of \textit{Haantjes coordinates}, allowing us to write simultaneously all $\boldsymbol{K}\in \mathscr{H}$ in a block-diagonal form. In particular, if $\mathscr{H}$ is an algebra of semisimple operators, then in Haantjes coordinates, all $\boldsymbol{K}\in \mathscr{H}$ acquire a diagonal form \cite{TT2021JGP}. 

\end{remark}

\vspace{2mm}

In the forthcoming considerations, given an operator $\boldsymbol{K}$, we shall denote by  $Spec(\boldsymbol{K}):= \{ \lambda_1(\bs{q}), \ldots, \lambda_s(\bs{q})\}$, $s\in \mathbb{N}\backslash\{0\}$,  the set of the pointwise distinct eigenvalues of  $\boldsymbol{K}$, assumed by default to be \emph{real}. The distribution of all the generalized eigenvector fields associated with the eigenvalue $\lambda_i=\lambda_i(\bs{q})$  will be denoted by
\begin{equation}
 \label{eq:DisL}
 \mathcal{D}_i: = \ker \Bigl(\boldsymbol{K}-\lambda_i\boldsymbol{I}\Bigr)^{\rho_i}, \qquad i=1,\ldots,s,
 \end{equation}
where $\rho_i \in \mathbb{N}\backslash \{0\}$ is the Riesz index of $\lambda_i$ (which by hypothesis will always be taken to be independent of $\bs{q}$). The value $\rho=1$ characterizes the \textit{proper eigen-distributions}, namely the eigen-distributions of proper eigenvector fields of $\bs{K}$.

\subsection{The symplectic-Haantjes manifolds}
Symplectic-Haantjes (or $\omega \mathscr{H}$) manifolds are  symplectic manifolds endowed with an  algebra of Haantjes operators compatible with the symplectic form. Apart from their mathematical properties,   these geometric structures are relevant because they provide a simple but sufficiently general setting in which we can naturally formulate the theory of Hamiltonian integrable systems.
\begin{definition}\label{def:oHman}
A symplectic-Haantjes (or $\omega \mathscr{H}$) manifold  of class $m$ is a triple $( M,\omega,\mathscr{H})$ which satisfies the following properties:
\begin{itemize}
\item[(i)]
$(M,\omega)$  is a   symplectic  manifold of dimension $ 2 \, n$;
\item[(ii)]
$\mathscr{H}$ is an Abelian Haantjes algebra of rank $m$;
\item[(iii)]
$(\omega,\mathscr{H})$ are algebraically compatible, that is
\beq
\omega(X,\boldsymbol{K} Y)=\omega(\boldsymbol{K} X,Y)  \qquad \forall \boldsymbol{K} \in \mathscr{H}.
\eeq
\end{itemize}
\end{definition}
\subsection{Haantjes chains}
The theory of Magri chains (also known as Lenard chains) is a fundamental piece of the geometric approach to soliton hierarchies. Magri chains have been introduced to construct integrals of motion in
involution for infinite-dimensional Hamiltonian systems. The related notion of Haantjes chains, defined in \cite{TT2022AMPA} and in \cite{TT2016SIGMA}, will be extensively used in the forthcoming analysis.
\begin{definition} 
 Let $( M,\mathscr{H})$ be a Haantjes algebra of rank $m$. We shall say that a smooth function $H$ generates a Haantjes chain of closed 1-forms of length $m$ if  there exists
a  basis  $\{\boldsymbol{K}_1,\ldots, \boldsymbol{K}_m\}$ of $\mathscr{H}$
 such that
\begin{equation} \label{eq:MHchain}
\rd (\boldsymbol{K}^T_\alpha \,\rd H )=\boldsymbol{0} , \quad\qquad \alpha=1,\ldots ,m ,
\end{equation}
where $\boldsymbol{K}^{T}_{\alpha}: \mathfrak{X}^{*}(M) \to \mathfrak{X}^{*}(M)$ is the transposed operator of $\boldsymbol{K}_{\alpha}$. The (locally) exact 1-forms 
\beq \label{eq:2.9}
\rd H_\alpha=\boldsymbol{K}^T_\alpha \,\rd H,
\eeq
which are supposed to be linearly independent, are the elements of the Haantjes chain of length $m$ generated by $H$. The functions $H_\alpha$ are said to be  the potential functions of the chain.
\end{definition}

\begin{lemma} \cite{TT2022AMPA} \label{lemma:cad_AMPA}
Let $(M, \omega, \mathscr{H})$ be an Abelian $\omega \mathscr{H}$ manifold. Then, the potential functions $H_{\alpha}$, whose differentials belong to all Haantjes chains generated by a single function $H$, are in involution among each other and with $H$, w.r.t. the Poisson bracket associated to the symplectic form $\omega$.
\end{lemma}

\subsection{Construction of DH coordinates}
We can diagonalize algebras of semisimple Haantjes operators by constructing a suitable coordinate chart. To this aim, we propose the following procedure \cite{TT2021JGP,TT2022AMPA,RTT2023JNS}.

\vspace{2mm}

1) Given an algebra $\mathscr{H}$, determine the nontrivial intersections $\mathcal{V}_a$ of the eigen-distributions of its operators (joint eigen-distributions). \\
2) For each of the annihilators $(\mathcal{V}_a)^{\circ}$,  construct a basis of closed one-forms. \\
3)  Find the characteristic coordinates by integrating them and imposing the canonical commutation relations. \\

The coordinate chart so determined represents a set of DH coordinates for the Haantjes operators of $\mathscr{H}$, namely they simultaneously diagonalize all the operators of $\mathscr{H}$ and are Darboux coordinates for the (compatible) symplectic form. 

\subsection{Separation of variables in the Haantjes geometry}
A relevant result is  Benenti's Theorem \cite{Ben80}, stating that
a family of Hamiltonian functions $(H_1,\ldots, H_n)$ are separable in a set of
canonical coordinates $(\boldsymbol{q},\boldsymbol{p})$ if and only if  they satisfy the relations
\begin{equation} \label{eq:SI}
\{H_i,H_j\}\lvert_k=\frac{\partial H_i}{\partial q^k}
\frac{\partial H_j}{\partial p_k}-\frac{\partial H_i}{\partial p_k}
\frac{\partial H_j}{\partial q^k}=0 \ , \quad 1\le k \le n
\end{equation}
where no summation over $k$ is understood. In this case, we shall say that the Hamiltonians $(H_1,\ldots, H_n)$ are in \textit{total separable  involution} (or $\mathcal{T}$-involution). 
The main results concerning the separability of a Hamiltonian system in the context of Haantjes geometry can be summarized as follows.

\begin{theorem} [Jacobi-Haantjes, \cite{RTT2022CNS}] \label{th:SoVgLc} 
Let $M$ be an Abelian  semisimple $\omega \mathscr{H}$ manifold of class $n$ and $\{H_1,H_2,\ldots,H_n\}$ be a set of $C^{\infty}(M)$ functions  belonging to a    Haantjes chain generated by a function $H\in C^{\infty}(M)$ via a basis of operators $\{\boldsymbol{K}_1,\ldots,\boldsymbol{K}_{n}\}\in \mathscr{H}$. Then, each set $(\bs{q},\bs{p})$  of {\rm DH} coordinates provides us with separation variables for the Hamilton--Jacobi equation associated with each function $H_j$.
 \par
 Conversely, if $M$ is a symplectic manifold and $\{H_1,H_2,\ldots,H_n\}$ are  $n$ independent,  $C^{\infty}(M)$ functions in total separable involution in a set of Darboux coordinates $(\bs{q},\bs{p})$, then they belong to the Haantjes  chain generated by the operators 
 \begin{equation} \label{eq:LSoV}
\boldsymbol{K}_\alpha=\sum _{i=1}^n \frac{\frac{\partial H_{\alpha}}{\partial p_i}}{ \frac{\partial H}{\partial p_i}}\bigg (\frac{\partial}{\partial q^i}\otimes \rd q^i +\frac{\partial}{\partial p_i}\otimes \rd p_i \bigg ),\qquad \alpha=1,\ldots,n ,
\end{equation}
where $H$ is any of the functions $\{H_1, \ldots, H_n\}$, with $\frac{\partial H}{\partial p_i}\neq 0$, $i=1,\ldots,n$. These operators generate a semisimple $\omega \mathscr{H}$ structure on $M$.
\end{theorem}

Notice that in the first implication of the previous theorem, the set of functions $\{H_1,H_2,\ldots,H_n\}$ is in involution by virtue of Lemma \ref{lemma:cad_AMPA}.

\begin{example}
Let us consider the Hamiltonian function \eqref{eq:1.1} and assume it is independent of the  coordinate $q^a$ (i.e., $q^a$ is ignorable). Then, as is well known, the conjugate momenta $p_a$ is an integral of motion, as it is in involution with $H$. More precisely, it is in total separable involution with $H$. In such a situation, we can write down a Haantjes operator $\bs{K}_a$ such that
$
\bs{K}_a^T \rd H=\rd p_a.
$
We obtain
\begin{equation}\label{eq:Km}
\bs{K}_a=\frac{1}{g^{ai}(p_i+A_i(\bs{q}))}
\left(\frac{\partial}{\partial q^a} \otimes \rd q^a +\frac{\partial}{\partial p_a} \otimes \rd p_a \right).
\end{equation}
In fact, from Eq. \eqref{eq:LSoV} it follows immediately
$$
\lambda _b^{(a)}=\frac{\frac{\partial H_a}{\partial p_b}}{\frac{\partial H}{\partial p_b}}=\frac{\delta_{ab}}{g^{bj}(p_j+A_j)} \ ,
$$
where $H_a:=p_a$.
\end{example}
We propose here a new result, which completes and generalizes a statement of \cite{RTT2022CNS} on \textit{multiseparable Hamiltonian systems}, i.e., systems admitting more than one set of separation coordinates. 
\begin{proposition} \label{th:multi}
An integrable Hamiltonian system possesses  as many inequivalent separation coordinate systems as the number of independent semisimple  $\omega\mathscr{H}$ manifolds of class $n$ that it admits. Also, for each semisimple  $\omega\mathscr{H}$ manifold of class $r<n$, there exists an associated set of coordinates assuring partial separation of the HJ equations admitted by the Hamiltonian system.
\end{proposition} 
This result is a direct consequence of Corollary 1 in \cite{RTT2022CNS} and of Theorem 5 in \cite{RTT2023prepr}.
\subsection{Generalized St\"ackel systems and symplectic-Haantjes manifolds} \label{subsec:2.5}
A large class of generalized St\"ackel systems was introduced by Arnold, Kozlov, and Neishtadt (AKN) in \cite{AKN1997}. As proved in \cite{TT2016SIGMA}, we can describe these systems in the framework of the symplectic-Haantjes geometry. Indeed, 
  let us consider the AKN Hamiltonian functions 
\begin{equation}  \label{eq:HStack}
H_j=\sum_{k=1}^n\frac{\tilde{S}_{jk}}{\textrm{det}(S)} f_k(q^k,p_k), \qquad j=1,\dots,n ,
  \end{equation}
where $S_{ij}$ are the elements of a St\"ackel matrix $S(\bs {q})$ (i.e., an invertible matrix whose $i$-th row depends  on the coordinate $q^i$ only), $\tilde{S}_{jk}$ denotes the cofactor of the element $S_{kj}$; $f_k(q^k,p_k)$ are called the \textit{St\"ackel functions}. Thus, the Haantjes chains
\begin{equation} \label{eq:HcS}
\boldsymbol{K}_{j} ^T\mathrm{d} H_1=\mathrm{d} H_{j}, \qquad j=1,\ldots,n
\end{equation}
admit as solutions the Haantjes operators defined by
\begin{equation} \label{eq:StackHaan}
\boldsymbol{K}_{j}:=\sum_{r=1}^n \frac{ \tilde{S}_{jr}}{ \tilde{S}_{1r}} \bigg( \frac{\partial}{\partial q^r}\otimes \mathrm{d} q^r+ \frac{\partial}{\partial p_r}\otimes \mathrm{d} p_r\bigg),
 \quad \ j=1,\ldots,n.
\end{equation}
\begin{remark}
The Haantjes operators \eqref{eq:StackHaan} are independent of the St\"ackel functions $f_k(q^k,p_k)$.
\end{remark}
An interesting property of the Haantjes operators associated with an AKN system is that they can be projected along the fibers of
$T^{*}Q$ by means of the canonical projection map
$\pi: T^{*} Q\rightarrow Q, (\bs{q},\bs{p})\mapsto \bs{q}$. We have the following
\begin{proposition} \label{pr:Kp}
The Haantjes operators \eqref{eq:StackHaan} can be projected along the fibers of $T^{*}Q$  onto the operators
\begin{equation}\label{eq:Kp}
\tilde{\boldsymbol{K}}_{j}:=\sum_{r=1}^n\frac{ \tilde{S}_{jr}}{ \tilde{S}_{1r}} \,  \frac{\partial}{\partial q^r}\otimes \mathrm{d} q^r, \quad \ j=1,\ldots,n-1.
\end{equation}
\end{proposition}
For the classical  St\"ackel systems, such operators represent Killing tensors, as we shall clarify below. 

\subsection{Classical  St\"ackel systems}
A particular, but important case is that of classical separable St\"ackel systems. They arise from Eq. \eqref{eq:HStack}
when the St\"ackel functions are quadratic  in the momenta:
\begin{equation}\label{eq:cSf}
f_k:=\frac{1}{2} p_k^2+W_k(q^k), \qquad k=1,\ldots,n.
\end{equation}
With this choice, we obtain the Hamiltonian function 
$$
H_1=\frac{1}{2} \sum_{j=1}^n g^{jj}(\bs{q})\,p_j^2+ V(\bs{q}),
$$
where the functions $g^{jj}(\bs{q})=\frac{ \tilde{S}_{1j}}{det(S)}$
can be regarded as  the diagonal  components of the inverse of a metric tensor $g$ over the configuration space $Q$:
\begin{equation} \label{eq:Sg}
\boldsymbol{G}:=\sum_{j=1}^n g^{jj} \frac{ \partial}{\partial q^j} \otimes \frac{ \partial}{\partial q^j}.
\end{equation}
Besides, $V(\bs{q})=\sum_{j=1}^n g^{jj} W_j$ is the corresponding potential energy.
Using metric \eqref{eq:Sg}, we can define the contravariant  form of the diagonal Haantjes operators
\eqref{eq:Kp}.
\begin{remark} It is also noteworthy that the tensor fields   \eqref{eq:Kp}  are Killing tensors for the  metric \eqref{eq:Sg} and are in involution with respect to the Schouten bracket of two symmetric contravariant tensors.
\end{remark} 

\subsection{Magnetic fields and Riemannian geometry} 
Let us recall some notions that we will use in what follows. 
\begin{definition}
Let $Q$ be a Riemannian manifold, and $\bs{G}$ be the inverse of its metric tensor. It defines the $C^\infty(Q)$-linear map (denoted, with a slight abuse of notation, by $\bs{G}$) and defined by $\bs{G}:\mathfrak{X}^\ast(Q)\to \mathfrak{X}(Q), \alpha \mapsto X=\bs{G}\,\alpha$ where $X$ is the unique vector field such that
$$ g(X,Y)=\langle \alpha, Y \rangle\qquad \forall \, Y\in \mathfrak{X}(Q) \ . $$
\end{definition}
Such isomorphism that maps  $1$-forms onto vector fields is also denoted as $X=(\alpha) ^\sharp$. The inverse of such isomorphism
$\bs{G}^{-1}:\mathfrak{X}(Q) \to \mathfrak{X}^\ast(Q), X \mapsto \alpha=\bs{G}^{-1}X$ is also  denoted as $\alpha=X^\flat$. 

Let us recall that, in a three-dimensional Riemaniann manifold, the Hodge star operator $\ast: \Lambda(Q)^k \to \Lambda(Q)^{3-k}$ evaluated on the basis two-forms gives us
\begin{equation}
 \ast(\rd q^i\wedge \rd q^j)=\iota_{\bs{G}\rd q^j}\iota_{\bs{G}\rd q^i} \sqrt{\det(g)}\,\rd q^1 \wedge\rd q^2\wedge \rd q^3,
\end{equation}
where $\iota_X$ denotes the interior product of a $k$-form with a vector field.

If $A$ denotes the vector field of the magnetic potential,
the  magnetic vector field can be defined as $B:=\nabla \times A=(\ast \, \rd(A)^\flat)^\sharp =\bs{G}\ast\rd(\bs{G}^{-1}A)$. In general curvilinear coordinates $(q^1,q^2,q^3)$, we can compute $B$ through the formula 
\begin{equation} \label{eq:rot}
\nabla \times A=\frac{1}{h_1 h_2 h_3}
\left|\begin{array}{ccc}
h_1\vec{e}_1& h_2\vec{e}_2 & h_3\vec{e}_3 \\
\frac{\partial}{\partial q^1} & \frac{\partial}{\partial q^2}  & \frac{\partial}{\partial q^3}  \\
A_1 & A_2 &A_3
\end{array}
\right|,
\end{equation}
where  $(\frac{\partial}{\partial q^1} ,\frac{\partial}{\partial q^2} ,\frac{\partial}{\partial q^3} )$ are the vector fields of the natural reference frame, $h_i=\sqrt{g_{ii}}$  their modules  and 
$(\vec{e}_1=\frac{1}{h_1}\frac{\partial}{\partial q^1} ,\vec{e}_2=\frac{1}{h_2}\frac{\partial}{\partial q^2} ,\vec{e}_3=\frac{1}{h_3}\frac{\partial}{\partial q^3})$ the unit vector fields  in the same directions. In the following, we shall denote by $(A_1,A_2,A_3)$ the components of a vector field $A(q^1,q^2,q^3)$ along the basis of the gradients 
$\left(\nabla q^1=\bs{G} \, \rd q^1, \nabla q^2=\bs{G} \, \rd q^2,\nabla q^3=\bs{G} \, \rd q^3\right)$.

\section{Separable systems with  magnetic field in cylindrical coordinates} \label{sec:3}

In this section, we shall determine the symplectic-Haantjes manifolds associated with a large class of magnetic systems in $\mathbb{E}^3$: those separating in cylindrical coordinates. Our strategy will be based on some natural ans\"atze together with the Jacobi-Haantjes Theorem \ref{th:SoVgLc}. 
\par

In cylindrical coordinates $(r,\varphi,z)$, the Euclidean metric reads $g= \text{diag}\,[1,r^2,1]$ and the Hamiltonian \eqref{eq:1.1} takes the form
\begin{equation} 
\begin{aligned}
H & = \dfrac{1}{2} \left[ (p_{r} + A_{r} (r, \varphi, z))^{2} + \frac{(p_{\varphi} + A_{\varphi} (r, \varphi, z))^{2}}{r^{2}} + (p_{z} + A_{z} (r, \varphi, z))^{2} \right] +V(r,\vphi,z) \\   & =\dfrac{1}{2} \bigg(\Pi_{r}^2+ \frac{\Pi_{\varphi}^{2}}{r^{2}} + \Pi_{z}^{2} \bigg) +V(r,\vphi,z), 
 \label{eq:phys_ham_woW_cyl}
\end{aligned}
\end{equation}
where $\Pi_{r} := p_{r} + A_{r} (r, \varphi, z)$, $\Pi_{\varphi} := p_{\varphi} + A_{\varphi} (r, \varphi, z)$ and $\Pi_{z} := p_{z} + A_{z} (r, \varphi, z)$ and $ (A_{r}, A_{\varphi} ,A_z)$ are respectively, the components of the linear momentum $\Pi=v$ and the vector magnetic potential $A$, along the reference frame $(\nabla r,\nabla \varphi,\nabla z)$. From Eq. \eqref{eq:rot} we obtain the expression of the magnetic field:
\begin{equation} \label{eq:3.2}
B(r,\varphi,z)=\frac{1}{r}\bigg(\frac{\partial A_z}{\partial \varphi} -\frac{\partial A_\varphi}{\partial z}\bigg) \, \vec{e}_r +
\bigg(\frac{\partial A_r}{\partial z}-\frac{\partial A_z}{\partial r}\bigg) \, \vec{e}_\varphi+
\frac{1}{r}
\bigg( \frac{\partial A_\varphi}{\partial r}-\frac{\partial A_r}{\partial \varphi}\bigg) \, \vec{e}_z,
\end{equation}
where $(\vec{e}_r, \vec{e}_\varphi, \vec{e}_z)$ are the unit vectors tangent to the coordinate lines. 
The Haantjes algebra $\mathscr{H}_{cyl}$, associated with the cylindrical web, is generated by the set of Haantjes (Nijenhuis)  operators
\begin{equation}
\bs{L}_1:=\left(\frac{\partial}{\partial r} \otimes \rd r +\frac{\partial}{\partial p_r} \otimes \rd p_r \right) \, ,
\end{equation}
\begin{equation}
\bs{L}_2:=\left(\frac{\partial}{\partial \varphi} \otimes \rd \varphi +\frac{\partial}{\partial p_\varphi} \otimes \rd p_\varphi \right) \, ,
\end{equation}
\begin{equation}
\bs{L}_3:=\left(\frac{\partial}{\partial z} \otimes \rd z +\frac{\partial}{\partial p_z} \otimes \rd p_z\right) \, .
\end{equation}
In the Cartesian frame $(x,y,z,p_x,p_y,p_z)$, they take the matrix form 
\beq \label{L}
\bs{L}_1=\frac{1}{x^2+y^2}\left[
\begin{array}{ccc|ccc}
x^2& xy & 0 & 0 & 0 & 0 \\
 xy &y^2 & 0 & 0 & 0 & 0 \\
 0 & 0 & 0 & 0 & 0 & 0 \\ \hline
 0 & x\,p_y-y\,p_x & 0& x^2 & xy & 0 \\
 -(x\,p_y-y\,p_x) & 0 & 0 & xy & y^2 & 0 \\
 0& 0 & 0 & 0 & 0 & 0 \\
\end{array}
\right] \, ,
\eeq
\begin{equation}
\bs{L}_2=\left[
\begin{array}{ccc|ccc}
 y^2 & -xy & 0 & 0 & 0 & 0 \\
 -xy &x^2 & 0 & 0 & 0 & 0 \\
 0 & 0 & 0 & 0 & 0 & 0 \\ \hline
 0 & -(x\,p_y-y\,p_x )& 0& y^2 & -xy & 0 \\
 x\,p_y-y\,p_x & 0 & 0 & -xy & x^2 & 0 \\
 0& 0 & 0 & 0 & 0 & 0 \\
\end{array}
\right]  \, ,
\end{equation}
\begin{equation}
\bs{L}_3= \text{diag}\, [0,0,1,0,0,1] \, . 
\end{equation}
Note that they can also be obtained using the complete Yano lift \cite{Y,IMM} of, respectively,  the evaluation of the Euler   tensor at the equatorial plane $z=0$ (up to the factor $(x^2+y^2)^{-1}$)
and the  inertia tensor (in $\mathbb{E}^2$) of a single particle \cite{TT2021JGP}.
\par
We obtained three cases of general systems in a magnetic field separable in cylindrical coordinates.

\vspace{2mm}

\textit{\underline{Case 1}}. Let us look for an integrable system that admits as integrals of motion the momenta conjugated to the coordinates   $(\varphi,z)$:
\begin{equation} \label{eq:Hcil1int}
H_2:= p_\varphi \, , \qquad H_3:=p_z \, .
\end{equation}
With this ansatz, the coordinates   $(\varphi,z)$ are ignorable in the Hamiltonian function \eqref{eq:phys_ham_woW_cyl}.
Therefore, 
\begin{equation}
 A(r)=A_r(r) \, \nabla r+ A_\varphi(r) \, \nabla \varphi+A_z(r) \, \nabla z  \, , \qquad V=V(r).
\end{equation}
From  Eq. \eqref{eq:3.2},   we obtain the magnetic field
\begin{equation} \label{eq:3.8}
B=-A_z^{\prime}(r) \, \vec{e}_\varphi+\frac{A_\varphi^{\prime}(r)}{r}\, \vec{e}_z\, ,
\end{equation}
which is tangent to the cylindrical surfaces of the cylindrical web. As it is independent of $A_r$, we can set $A_r=0$ without loss of generality.
In view of Eq. \eqref{eq:phys_ham_woW_cyl}, the Hamiltonian function of this system turns out to be
\begin{equation}\label{eq:cil1}
H=\frac{1}{2}\bigg(p_r^2+\frac{p_\varphi^2}{r^2} +p_z^2 \bigg) +\frac{A_\varphi(r)}{r^2} \, p_\varphi +A_z(r) \, p_z+V(r)+\frac{1}{2}\bigg(\frac{A_\varphi^2(r)}{r^2}+A_z^2(r)\bigg)\, .
\end{equation}
Thus, the Hamiltonian functions \eqref{eq:Hcil1int}, \eqref{eq:cil1} are not only in involution but even in total separable involution. Then,  Eq. \eqref{eq:Km}  implies that the corresponding semisimple Haantjes operators that solve the chain equations $\boldsymbol{K}_{2}^{T} \rd H = \rd H_{2}$ and $\boldsymbol{K}_{3}^{T} \rd H = \rd H_{3}$ are
\begin{equation}
 \boldsymbol{K}_{2} =  \frac{r^2}{p_{\varphi} +  A_\varphi(r)}\,\bs{L}_2, \qquad
 \boldsymbol{K}_{3} = \frac{1}{p_z+A_z(r)} \, \bs{L}_3.
\end{equation}
Therefore, the Hamilton-Jacobi equations associated to it are
\begin{equation*}
\begin{cases}
\frac{1}{2}\bigg(\left( \frac{\partial W}{\partial r} \right)^2+\frac{h_2^2}{r^2} +h_3^2 \bigg) +\frac{A_\varphi(r)}{r^2} \, h_2 +A_z(r) \, h_3 +V(r)+\frac{1}{2}\bigg(\frac{A_\varphi^2(r)}{r^2}+A_z^2(r)\bigg) = h_1 \, , \\
\frac{\partial W}{\partial \varphi} = h_2 \, , \\
\frac{\partial W}{\partial z} = h_3 \, .
\end{cases}
\end{equation*}
By putting $p_r=\frac{\partial W}{\partial r}$,  $p_z=\frac{\partial W}{\partial z}$, $p_{\varphi}=\frac{\partial W}{\partial \varphi}$ in the previous formulas, one obtains the Jacobi-Skyanin separation relations.

Since it is separated, we can split $W$ as $W (r, \varphi, z) = W_1 (r) + W_2 (\varphi) + W_3 (z)$. Thus, the resulting system consists of three ODEs, and solving them yields
\begin{equation}
W(r,\varphi,z;h_1,h_2,h_3)=h_2 \, \varphi+h_3 \, z \pm 
\int \sqrt{-2 \left( \frac{h_2^2}{2 r^2}+\frac{h_3^2}{2}-h_1+\frac{A_\varphi (r)}{r^2}h_2 +A_z(r) \, h_3+U(r) \right)}\, \rd r \ ,
\end{equation}
where $U(r)= V(r)+\frac{1}{2}\bigg(\frac{A_\varphi^2(r)}{r^2}+A_z^2(r)\bigg)$, $h_1,h_2,h_3$ are the values taken by the integrals of motion $(H,H_2,H_3)$ on the leaves of the Lagrangian foliation.
\vspace{2mm}

\textit{\underline{Case 2}}.
Let us seek an integrable system possessing the momentum conjugated  to  the coordinate $\vphi$ as an integral of motion:
\begin{equation} \label{eq:Hcil2int}
H_2:= p_\varphi.
\end{equation} 
With this ansatz, the coordinates   $\varphi$ must be ignorable in the Hamiltonian \eqref{eq:phys_ham_woW_cyl}.
This implies  that  the vector  and the scalar potentials are, respectively, of the form
\begin{equation}
A(r,z)=A_r (r,z) \, \nabla r+A_\varphi(r,z) \, \nabla \varphi+A_z(r,z) \, \nabla z \, , \qquad  V=V(r,z) \, , 
\end{equation}
where the functions $A_r, A_\varphi, A_z, V$ are smooth arbitrary  functions  of their arguments. We deduce for the magnetic field the expression 
\begin{equation}
B=-\frac{1}{r}\frac{\partial A_\varphi}{\partial z} \, \vec{e}_r + \bigg(  \frac{\partial A_r}{\partial z}-\frac{\partial A_z}{\partial r} \bigg) \,  \vec{e}_\varphi +\frac{1}{r}
 \frac{\partial A_\varphi}{\partial r} \,  \vec{e}_z \, , 
\end{equation}
so that the Hamiltonian function turns out to be
\begin{equation} \label{eq:Hcil2}
\begin{split}
H=\frac{1}{2} \bigg( p_r^2+\frac{p_\varphi^2}{r^2} +p_z^2 \bigg) +A_r(r,z) \, p_r+\frac{A_\varphi(r,z)}{r^2} \, p_\varphi+A_z(r,z) \, p_z & \\ +V(r,z)+\frac{1}{2} \bigg( A_r^2(r,z)+\frac{A_\varphi^2(r,z)}{r^2}+A_z^2(r,z) \bigg), &
\end{split}
 \end{equation}
which is in separable involution with the integral \eqref{eq:Hcil2int}.
From Eq. \eqref{eq:Km} we get the semisimple Haantjes operator for the chain $\boldsymbol{K}_{2}^{T} \rd H = \rd H_{2}$:
\begin{equation}
 \boldsymbol{K}_{2} = \frac{r^2}{p_\varphi+A_\varphi(r,z)} \, \bs{L}_2 \, .
\end{equation}
In order to determine another integral of motion in separable involution with \eqref{eq:Hcil2} and \eqref{eq:Hcil2int}  and a corresponding Haantjes operator $\boldsymbol{K}_{3}$, we can assume that $\varphi$ is ignorable in $\boldsymbol{K}_{3}$ and that, for simplicity, it depends on $p_\varphi, p_z$ but not on $p_r$:
\[
\boldsymbol{K}_{3}= \text{diag} \, [a_1,a_2,a_3,a_1,a_2,a_3] \, ,
\]
where $a_i=a_i(r,z,p_\varphi,p_z)$, $i=1,2,3$.
By requiring the local exactness condition $\rd (\boldsymbol{K}_{3}^T \rd H)=\bs{0}$, we get an overdetermined system of six PDEs in three unknowns. We obtain the particular solution
\begin{equation}
A(r,z)= 
\bigg(f_2(r)+ r^2 \, f_{3} (z)\bigg) \,  \nabla \varphi
, \qquad
B=-rf_3^{\prime}(z)\vec{e}_r+\bigg(\frac{f_2^{\prime}(r)}{r}+2 f_3(z)\bigg)\vec{e}_z
\end{equation}
\begin{equation}
\mathit{V}  = 
f_{5} (r)+f_{4} (z)-\frac{r^{2} f_{3}^2 (z)}{2}-f_{2} (r) f_{3} (z),
\end{equation}
\begin{equation}
\boldsymbol{K}_{3}= 
\frac{r^{2} \, f_{3} (z)}{ p_{\vphi} +A_\varphi} \, \bs{L}_2+\bs{L}_3 \, ,
\end{equation}
\begin{equation}
H_3=\frac{1}{2} \, p_z^2 +f_3(z) \, p_\varphi+f_4(z),
\end{equation} 
where $f_2$, $f_3$, $f_4$, $f_5$ are arbitrary functions. 


The complete integral of the set of Hamilton-Jacobi equations is 
\begin{equation}
\begin{split}
W(r,\varphi,z;h_1,h_2,h_3)=h_2 \varphi 
& \pm\int \sqrt{-2 \left( \frac{h_2^2}{2 r^2} - h_1+\frac{f_2(r)}{r^2}h_2 +h_3+f_6(r) \right)}\, \rd r  \\
& \pm\int \sqrt{-2 \left( f_3(z)h_2 -h_3+f_4(z)  \right)}\, \rd z \ ,
\end{split}
\end{equation}
where $f_6(r)=f_5(r)+f_2(r)^2/2r^2$, $h_1,h_2,h_3$ are the values taken by the integrals of motion $(H,H_2,H_3)$ on the leaves of the Lagrangian foliation.

\vspace{2mm}

\textit{\underline{Case 3}}. Let us proceed as in Case 2, by interchanging $\varphi$ with $z$. To this aim, let us look for a Hamiltonian system    admitting the integral 
\begin{equation} \label{eq:Hcil3int}
H_3:= p_z \, .
\end{equation}
With this ansatz, the coordinate   $z$ is ignorable in the Hamiltonian function \eqref{eq:phys_ham_woW_cyl}.
This implies  that the vector  and the scalar potential are of the form
\begin{equation}
A(r,\varphi)=A_r (r,\varphi) \, \nabla r +A_\varphi(r,\varphi) \, \nabla \varphi+A_z(r,\varphi) \, \nabla z \, , \qquad  V=V(r,\varphi),
\end{equation}
where the functions $A_r, A_\varphi, A_z, V$ are arbitrary (smooth) functions of $r$ and periodic functions of $\vphi$. 
The associated magnetic field is
\begin{equation}
B=\frac{1}{r}\frac{\partial A_z}{\partial \varphi} \, \vec{e}_r-\frac{\partial A_z}{\partial r} \, \vec{e}_\varphi +
\frac{1}{r}\bigg( \frac{\partial A_\varphi}{\partial r}-\frac{\partial A_r}{\partial \varphi} \bigg) \, \vec{e}_z \, .
\end{equation}

The Hamiltonian function of this system reads
\begin{equation} \label{eq:Hcil3}
\begin{split}
H = \frac{1}{2} \bigg( p_r^2+\frac{p_\varphi^2}{r^2} +p_z^2 \bigg) +A_r(r,\varphi) \, p_r+\frac{A_\varphi(r,\varphi)}{r^2} \, p_\varphi + A_z(r,\varphi) \, p_z & \\  + V(r,\varphi)+\frac{1}{2}\left(A_r^2(r,\varphi)+\frac{A_\varphi^2(r,\varphi)}{r^2}+A_z^2(r,\varphi)\right) \, , &
\end{split}
 \end{equation}
 that is in separable involution with the integral \eqref{eq:Hcil3int}. From Eq. \eqref{eq:Km} we get that the following semisimple Haantjes operator solves the chain equations $\boldsymbol{K}_{3}^{T} \rd H = \rd H_{3}$:
\begin{equation}
 \boldsymbol{K}_{3} = \frac{1}{p_z+A_z(r,\varphi)} \, \bs{L}_3 \, .
\end{equation}
By analogy with the previous case, we can determine another integral of motion in separable involution with \eqref{eq:Hcil3} and \eqref{eq:Hcil3int}, and the associated Haantjes operator $\boldsymbol{K}_{2}$ by assuming that $z$ is ignorable in $\boldsymbol{K}_{2}$ and that it depends on $p_\varphi, \, p_z$ but not on $p_r$:
\[
\boldsymbol{K}_{2}= \text{diag} \, [b_1,b_2,b_3,b_1,b_2,b_3],
\]
where $b_i=b_i(r,\varphi,p_\varphi,p_z)$, $i=1,2,3$.
The local exactness condition $\rd (\boldsymbol{K}_{2}^T \rd H)=\bs{0}$ provides us with an overdetermined system of PDEs, from which we obtain
\begin{equation}
A= 
\bigg(f_{1} (r)+\frac{f_{2}(\varphi)}{r^{2}}\bigg) \,  \nabla z \, ,\hspace{2mm}
B=\frac{f_2^{\prime}(\varphi)}{r^3} \vec{e}_r+
\bigg(-f_1^{\prime}(r)+2\frac{f_2(\vphi)}{r^3}\bigg)\vec{e}_\varphi \, ,
\end{equation}
\begin{equation}
 V =  f_3(r)
-\frac{{f_{1}}(r) f_{2} (\varphi)}{r^2}-\frac{f_2^2(\varphi)}{2 r^4}+\frac{f_4(\varphi)}{r^2} \, ,
\end{equation}
\begin{equation}
\boldsymbol{K}_{2} = r^2 \, \bs{L}_2+\dfrac{f_2(\varphi)}{p_z+A_z}\bs{L}_3 \, ,
\end{equation}
\begin{eqnarray}
H_2&=&\frac{p_{\varphi}^{2}}{2}+ f_{2} (\varphi) \, p_{\mathit{z}} +f_4(\varphi) \, .
\end{eqnarray}
The complete integral of the set of Hamilton-Jacobi equations is 
\begin{equation}
\begin{split}
W(r,\varphi,z;h_1,h_2,h_3)=h_3 \, z  
& \pm\int \sqrt{-2 \left( \frac{h_3^2}{2} - h_1+\frac{h_2}{r^2}+ f_1(r)h_3+\frac{1}{2} f_1^2(r)+f_3(r) \right)}\, \rd r  \\
& \pm\int \sqrt{-2 \left(-h_2+ f_2(\varphi)h_3 +f_4(\varphi)  \right)}\, \rd \varphi \ .
\end{split}
\end{equation}
\begin{remark}
Let us note that, if we trade the triple of the Hamiltonian functions $(H_1,H_2,H_3)$ with the triple $(\tilde{H}_1:=H_1-\frac{1}{2} H_3^2, \tilde{H}_2:=H_2,\tilde{H}_3:=H_3)$, we get a St\"{a}ckel system with the following St\"ackel matrix and St\"ackel vector 
$$
S(\bs{q})=
\left[\begin{array}{ccc}
1 & -\frac{1}{r^2} & -f_1(r) \\
0 & 1 & -f_2(\varphi) \\
0 & 0 & 1
\end{array}\right] \ ,
\qquad
F(\bs{q},\bs{p})=\left[
\begin{array}{c}
\frac{p_r^2}{2}+\frac{f_1^2(r)}{2} +f_3(r) \\
\frac{p_\varphi^2}{2}+f_4(\varphi)\\
p_z 
\end{array}
\right]
\ .
$$
The modified Hamiltonian functions $\tilde{H}_1,\tilde{H}_2, \tilde{H}_3$ form two Haantjes chains by means of the modified Haantjes operators, constructed via Eq. \eqref{eq:StackHaan}:
$$
\tilde{\bs{K}}_2=r^2 \bs{L}_2+\frac{f_2(\varphi)}{A_z} \bs{L}_3\ , \qquad \tilde{\bs{K}}_3=\frac{1}{A_z} \bs{L}_3;
$$
 however, they belong to the same Haantjes algebra $\mathscr{H}_{\text{cyl}}$. Thus, we can say that the Lagrangian foliation generated by $(H_1,H_2,H_3)$ admits the St\"ackel basis $(\tilde{H}_1,\tilde{H}_2, \tilde{H}_3)$. We have not been able to find the St\"ackel basis for Cases 1 and 2.  The general problem of constructing a St\"ackel basis for a given Lagrangian foliation will be addressed elsewhere.
\end{remark}

In  \cite{BCRJMP2001}, it has been proved that the  three cases discussed above give  the most general  systems in a  magnetic field separable in cylindrical coordinates. Thus, our analysis in the geometric setting of Haantjes algebras is coherent with the results obtained in \cite{BCRJMP2001} with a different approach.

\section{Superintegrability and multiseparability} \label{sec:4}

The purpose of this section is to construct the symplectic-Haantjes manifolds associated with several cases of physically relevant superintegrable magnetic systems. 
According to Corollary \ref{th:multi}, we shall determine as many Haantjes structures as the number of separation variables admitted by the system under analysis.

We shall first consider the case of the constant magnetic field. 
On the one hand, this case can be regarded as a specialization of the general discussion concerning separability in cylindrical coordinates and Haantjes geometry developed in Section \ref{sec:3}.
On the other hand, the presence of a constant magnetic field allows the system to possess two additional integrals of motion.
This, in turn, implies that there exist additional symplectic-Haantjes structures related to the extra integrals, providing total or partial separation coordinates \cite{RTT2023prepr}.
\begin{remark}
In Cartesian coordinates, the physical Hamiltonian that describes a system defined in a Euclidean 3D space with a magnetic field and vanishing potential is
\begin{equation} \label{eq:phys_ham}
H = \dfrac{1}{2} \big( \bs{p} + A (\bs{x}) \big)^{2},
\end{equation}
with $\bs{x}=(x,y,z)$.  The kinetic momentum of the system, which coincides with the velocity vector in our units, is given by \cite{Jackbook}
\begin{equation}
\Pi(\bs{x}) := \bs{p} + A (\bs{x}) \, .
\end{equation}

\end{remark}

\subsection{A superintegrable model in a constant magnetic field}

The simplest magnetic system we can study in three dimensions is the one in a constant magnetic field, that  without loss of generality we can define as
\begin{equation}
B (x,y,z) = b ~\vec{e}_{z} \, ,
\end{equation} 
where $b\in \mathbb{R}$ is an arbitrary constant. The Hamiltonian model \eqref{eq:phys_ham} is maximally superintegrable \cite{Marchesiello2015}. The integrals of the motion of first degree in the momenta are
\beq
\begin{aligned}
H_{1} & = \Pi_{x} + b \ y \, , \\
H_{2} & = \Pi_{y} - b \ x \, , \\
H_{3} & = \Pi_{z} \, , \\
H_{4} & = \Lambda_{z} - \dfrac{b}{2} (x^{2} + y^{2}) \, .
\end{aligned}
\eeq
Here $\Lambda_{z} = x\, \Pi_{y} - y\, \Pi_{x}$ is the z-component of the total angular momentum $\Lambda := \bs{x} \times \Pi$.
The set of these integrals jointly with the Hamiltonian \eqref{eq:phys_ham} is not functionally independent; however, another independent, nonpolynomial integral is given by
\begin{equation}
	H_{5} = -\Pi_x\cos\Big(\frac{b \, z}{\Pi_z}\Big)-\Pi_y\sin\Big(\frac{b \, z}{\Pi_z}\Big) \, .
\end{equation}
It commutes with $H_1$ and $H_2$. In the subsequent discussion, we shall construct Haantjes manifolds associated with commutative Poisson subalgebras generated by the previous integrals.

\subsubsection{Semisimple Haantjes operators and spectral analysis}
In the following, we shall provide a set of \textit{semisimple} Haantjes operators that solve the chain equations $\boldsymbol{K}_{i}^{T} \rd H = \rd H_{i}$, jointly with their spectral properties. We obtain for the first chain the following operator and the corresponding eigen-distributions: 
\begin{equation} \label{eq:4.7}
\boldsymbol{K}_{1} = \dfrac{1}{\Pi_{x}} \bs{R}_1,
\eeq
with
\beq \label{eq:R1}
\bs{R}_1:=\left[
\begin{array}{ccc|ccc}
 1 & 0 & 0 & 0 & 0 & 0 \\
 0 & 0 & 0 & 0 & 0 & 0 \\
 0 & 0 & 0 & 0 & 0 & 0 \\ \hline
 0 & \partial_{x} A_{y} & \partial_{x} A_{z} & 1 & 0 & 0 \\
 - \partial_{x} A_{y} & 0 & 0 & 0 & 0 & 0 \\
 - \partial_{x} A_{z} & 0 & 0 & 0 & 0 & 0 \\
\end{array}
\right] \, , 
\end{equation}
and
\begin{align*}
& \lambda_{1}^{(1)} = 0 \, , \quad \rho_{1}^{(1)} = 1 \, , \   \mathcal{D}_{1} = \bigg\langle \partial_{x} A_{z} \frac{\partial}{\partial y} - \partial_{x} A_{y} \frac{\partial}{\partial z}, \frac{\partial}{\partial y} - \partial_{x} A_{y} \frac{\partial}{\partial p_{x}}, \frac{\partial}{\partial p_{y}}, \frac{\partial}{\partial p_{z}} \bigg\rangle \, , \\
 & \lambda_{2}^{(1)} = \frac{1}{\Pi_{x}} \, , \quad \rho_{2}^{(1)} = 1 \, , \ \mathcal{D}_{2} = \bigg\langle \frac{\partial}{\partial x} - \partial_{x} A_{y} \frac{\partial}{\partial p_{y}} - \partial_{x} A_{z} \frac{\partial}{\partial p_{z}}, \frac{\partial}{\partial p_{x}} \bigg\rangle \, .
\end{align*}
For the second chain, we have
\begin{equation}  \label{eq:4.10}
\boldsymbol{K}_{2} = \dfrac{1}{\Pi_{y}} \bs{R}_2,
\eeq
with
\beq \label{eq:R2}
\bs{R}_2:=\left[
\begin{array}{ccc|ccc}
 0 & 0 & 0 & 0 & 0 & 0 \\
 0 & 1 & 0 & 0 & 0 & 0 \\
 0 & 0 & 0 & 0 & 0 & 0 \\ \hline
 0 & - \partial_{y} A_{x} & 0 & 0 & 0 & 0 \\
 \partial_{y} A_{x} & 0 & \partial_{y} A_{z} & 0 & 1 & 0 \\
 0 &  - \partial_{y} A_{z} & 0 & 0 & 0 & 0 \\
\end{array}
\right] \, ,
\end{equation}
and
\begin{align*}
& \lambda_{1}^{(2)} = 0 \, , \quad \rho_{1}^{(2)} = 1 \, , \  \mathcal{D}_{3} = \bigg\langle  \partial_{y} A_{z} \frac{\partial}{\partial x} - \partial_{y} A_{x} \frac{\partial}{\partial z}, \frac{\partial}{\partial x} - \partial_{y} A_{x} \frac{\partial}{\partial p_{y}}, \frac{\partial}{\partial p_{x}}, \frac{\partial}{\partial p_{z}} \bigg\rangle \, , \\
& \lambda_{2}^{(2)} = \frac{1}{\Pi_{y}} \, , \quad \rho_{2}^{(2)} = 1 \, , \  \mathcal{D}_{4} = \bigg\langle \frac{\partial}{\partial y} - \partial_{y} A_{x} \frac{\partial}{\partial p_{x}} - \partial_{y} A_{z}\frac{\partial}{\partial p_{z}}, \frac{\partial}{\partial p_{y}} \bigg\rangle \, .
\end{align*}
Solving the third chain, we obtain
\begin{equation} \label{eq:op_k3_bconst3d_cart}
\boldsymbol{K}_{3} = \dfrac{1}{\Pi_{z}} \bs{R}_3,
\eeq
with
\beq \label{eq:R3}
\bs{R}_3:= \left[
\begin{array}{ccc|ccc}
 0 & 0 & 0 & 0 & 0 & 0 \\
 0 & 0 & 0 & 0 & 0 & 0 \\
 0 & 0 & 1 & 0 & 0 & 0 \\ \hline
 0 & 0 & - \partial_{z} A_{x} & 0 & 0 & 0 \\
 0 & 0 & -\partial_{z} A_{y} & 0 & 0 & 0 \\
\partial_{z} A_{x} & \partial_{z} A_{y} & 0 & 0 & 0 & 1 \\
\end{array}
\right] \, , 
\end{equation}
\begin{align*}
& \lambda_{1}^{(3)} = 0 \, , \quad \rho_{1}^{(3)} = 1 \, , \  \mathcal{D}_{5} = \bigg\langle \partial_{z} A_{y} \frac{\partial}{\partial x} - \partial_{z} A_{x} \frac{\partial}{\partial y}, \frac{\partial}{\partial x} - \partial_{z} A_{x} \frac{\partial}{\partial p_{z}}, \frac{\partial}{\partial p_{x}}, \frac{\partial}{\partial p_{y}} \bigg\rangle \, , \\
& \lambda_{2}^{(3)} = \frac{1}{\Pi_{z}} \, , \quad \rho_{2}^{(3)} = 1 \, , \  \mathcal{D}_{6} = \bigg\langle \frac{\partial}{\partial z} - \partial_{z} A_{x} \frac{\partial}{\partial p_{x}} - \partial_{z} A_{y} \frac{\partial}{\partial p_{y}}, \frac{\partial}{\partial p_{z}} \bigg\rangle \, .
\end{align*}
The operators $\bs{R}_1$, $\bs{R}_2$, $\bs{R}_3$ introduced above will play a significant role in the discussion of Section \ref{sec:6}. Finally, 
\begin{equation}
\boldsymbol{K}_{4} = \dfrac{1}{\Lambda_{z}}
\left[
\begin{array}{ccc|ccc}
 y^{2} & - x y & 0 & 0 & 0 & 0 \\
 - x y & x^{2} & 0 & 0 & 0 & 0 \\
 0 & 0 & 0 & 0 & 0 & 0 \\ \hline
 0 & f_{4} & g_{4} & y^{2} & - x y & 0 \\
 - f_{4} & 0 & h_{4} & - x y & x^{2} & 0 \\
 - g_{4} & - h_{4} & 0 & 0 & 0 & 0 \\
\end{array}
\right] \, , \label{eq:op_k4_bconst3d_cart}
\end{equation}
where
\begin{equation}
\begin{aligned}
f_{4} & = - \Lambda_{z} + x y \, ( \partial_{x} A_{x} - \partial_{y} A_{y} ) - x^2 \, \partial_{y} A_{x} + y^2 \, \partial_{x} A_{y} \, , \\
g_{4} & = y  \, \left( - x \ \partial_{y} A_{z} + y \ \partial_{x} A_{z} \right) \, , \\
h_{4} & = x \, \left( x \ \partial_{y} A_{z} - y \ \partial_{x} A_{z} \right) \, .
\end{aligned}
\end{equation}
The spectral analysis for $\bs{K}_4$ can be performed in full analogy with the previous cases and is not reported here.

\subsubsection{Construction of the DH coordinates} 
The operators $\{\bs{K}_1,\ldots,\bs{K}_4\}$ generate several commutative Haantjes algebras $\mathscr{H}_i$. We shall determine the Darboux-Haantjes coordinates for these algebras, which both diagonalize all semisimple Haantjes operators of each algebra and leave the symplectic form invariant. For the sake of clarity, we discuss in full detail the case of the algebra  $\mathscr{H}_{1} = \{ \boldsymbol{I}, \boldsymbol{K}_{1}, \boldsymbol{K}_{3} \}$. Defining the distributions
\begin{equation}
\mathcal{V}_{1} = \mathcal{D}_{1} \cap \mathcal{D}_{5}, \quad \mathcal{V}_{2} = \mathcal{D}_{1} \cap \mathcal{D}_{6}, \quad \mathcal{V}_{3} = \mathcal{D}_{2} \cap \mathcal{D}_{5}, \quad 
\end{equation}
we can work out the annihilators
\begin{equation}
\begin{aligned}
& (\mathcal{V}_{1} \oplus \mathcal{V}_{2})^{\circ} = \langle \rd x \, , \partial_{x} A_{y} \, \rd y + \partial_{z} A_{x} \, \rd z + \rd p_{x} \rangle, \\
& (\mathcal{V}_{1} \oplus \mathcal{V}_{3})^{\circ} = \langle \rd z \, , \partial_{x} A_{z} \, \rd x + \partial_{y} A_{z} \, \rd y + \rd p_{z} \rangle, \\
& (\mathcal{V}_{2} \oplus \mathcal{V}_{3})^{\circ} = \langle \rd y \, , \partial_{x} A_{y} \, \rd x + \partial_{z} A_{y} \, \rd z + \rd p_{y} \rangle \, .
\end{aligned} \label{eq:annih_h1_bconst3d_cart}
\end{equation}
Integrating them, we find the set of Darboux-Haantjes coordinates
\begin{equation} 
\begin{aligned}
& q^{1} = x \, , \qquad p_{1} = \Pi_{x} + b \, y \, , \\
& q^{2} = y \, , \qquad p_{2} = \Pi_{y} \, , \\
& q^{3} = z \, , \qquad p_{3} = \Pi_{z} \, .
\end{aligned} \label{eq:4.19}
\end{equation}
In this chart, the operators of the algebra $\mathscr{H}_{1}$ read
\begin{equation}
 \boldsymbol{K}_{1} =  \frac{1}{p_{1} - b \, q^{2}} \, \text{diag} \, [1, 0, 0, 1, 0, 0], \qquad
 \boldsymbol{K}_{3} = \frac{1}{p_{3}} \, \text{diag} \, [0, 0, 1, 0, 0, 1] \, ;
\label{eq:ops_h1_bconst3d_cart}
\end{equation}
and the Hamiltonian function and the integrals of the motion read
\beq \label{eq:4.22}
\begin{aligned}
& H = \frac{1}{2} \, \left[ (p_{1} - b \, q^{2})^{2} + p_{2}^{2} + p_{3}^{2} \right] \, , \\
& H_{1} = p_{1} \, , \\
& H_{3} = p_{3} \, . \\
\end{aligned}
\eeq
We shall similarly analyze the remaining algebras, skipping most technical details.  For the algebra  $\mathscr{H}_{2} = \{ \boldsymbol{I}, \boldsymbol{K}_{2}, \boldsymbol{K}_{3} \}$, we define the distributions
\begin{equation}
\mathcal{V}_{4} = \mathcal{D}_{3} \cap \mathcal{D}_{5} \, , \quad \mathcal{V}_{5} = \mathcal{D}_{3} \cap \mathcal{D}_{6} \, , \quad \mathcal{V}_{6} = \mathcal{D}_{4} \cap \mathcal{D}_{5} \,.
\end{equation}
Integrating the corresponding annihilators, we find the set of Darboux-Haantjes coordinates
\begin{equation}
\begin{aligned}
& q^{1} = x \, , \qquad p_{1} = \Pi_{x} \, , \\
& q^{2} = y \, , \qquad p_{2} = \Pi_{y} - b \, x \, , \\
& q^{3} = z \, , \qquad p_{3} = \Pi_{z} \, .
\end{aligned} \label{eq:coord_diag_h2_bconst3d_cart}
\end{equation}
In this chart, the operators of the algebra $\mathscr{H}_{2}$ read
\begin{equation}
 \boldsymbol{K}_{2} =  \frac{1}{p_{2} + b \, q^{1}} \, \text{diag} \, [0, 1, 0, 0, 1, 0] \, , \qquad
 \boldsymbol{K}_{3} = \frac{1}{p_{3}} \, \text{diag} \, [0, 0, 1, 0, 0, 1] \, ;
\label{eq:ops_h2_bconst3d_cart}
\end{equation}
and the Hamiltonian function and the integrals of the motion read
\beq
\begin{aligned}
& H = \frac{1}{2} \, \left[ p_{1}^{2} + (p_{2} + b \, q^{1})^{2} + p_{3}^{2} \right] \, , \\
& H_{2} = p_{2} \, , \\
& H_{3} = p_{3} \, . \\
\end{aligned}
\eeq
Finally, for the algebra  $\mathscr{H}_{3} = \{ \boldsymbol{I}, \boldsymbol{K}_{3}, \boldsymbol{K}_{4} \}$, the corresponding annihilators  can be easily integrated  in several ways, providing us with different choices for the DH coordinates. The most natural choice is that of cylindrical coordinates:
\begin{equation}
\begin{aligned}
& q^{1} = \sqrt{ x^{2} + y^{2} } \, , \\
& q^{2} = \arctan \left( \frac{y}{x} \right) \, , \\
& q^{3} = z \, ,
\end{aligned} \qquad
\begin{aligned}
& p_{1} = \dfrac{x \ \Pi_{x} + y \ \Pi_{y}}{\sqrt{ x^{2} + y^{2} }} \, , \\
& p_{2} = x \Big( \Pi_{y} - \frac{b}{2} x \Big) - y \Big( \Pi_{x} + \frac{b}{2} y \Big) \, , \\
& p_{3} = \Pi_{z} \, ;
\end{aligned} \label{eq:coord_diag_h3v3_bconst3d_cart}
\end{equation}
where the operators $\boldsymbol{K}_{3}$ and $\boldsymbol{K}_{4}$ take the form
\begin{equation} \label{eq:4.28}
 \boldsymbol{K}_{3} = \dfrac{1}{p_{3}} \bs{L}_3 \, , \qquad
 \boldsymbol{K}_{4} = \dfrac{2 (q^{1})^{2}}{2 \, p_{2} + b \ (q^{1})^{2}} \bs{L}_2 \, . 
\end{equation}

In this chart, the Hamiltonian function and the integrals of the motion read
\beq
\begin{aligned}
& H = \frac{1}{2} \Big[ p_{1}^{2} + \dfrac{1}{(q^{1})^{2}} \, \Big(p_{2} + \frac{b}{2} \, (q^{1})^{2} \Big)^{2} + p_{3}^{2} \Big] \, , \\
& H_{3} = p_{3} \, , \\
& H_{4} = p_{2} \, . 
\end{aligned}
\eeq
This case can also be obtained by specializing Case 1 of Section \ref{sec:3} by choosing $B=const.$ in Eq. \eqref{eq:3.8}.
\subsection{The helical undulator}

It is an interesting magnetic model \cite{Marchesiello2015}, described by the Hamiltonian  \eqref{eq:phys_ham} with a magnetic field of the form 
\begin{equation} \label{eq:4.52}
B (x,y,z) = - b_{3} \cos \Big( \frac{2 z}{a} \Big) \, \vec{e}_{x}  + b_{3} \sin \Big( \frac{2 z}{a} \Big) \, \vec{e}_{y} \, ,
\end{equation}
where $b_{3}$ and $a$ are arbitrary constants.\footnote{A slightly more general system can be obtained by adding a constant contribution along the $z$-axis to the magnetic field \eqref{eq:4.52}. The system so obtained is still superintegrable, but not separable in any orthogonal coordinate system on $Q$ (see \cite{KMSAOP2023}).}
The model is minimally superintegrable and the associated integrals of the motion are
\beq
\begin{aligned}
H_{1} & = \Pi_{x}  + \frac{b_{3} \, a}{2} \, \cos \Big( \frac{2 z}{a} \Big) \, , \\
H_{2} & = \Pi_{y} - \frac{b_{3} \, a}{2} \, \sin \Big( \frac{2 z}{a} \Big) \, , \\
H_{3} & = \Lambda_{z} - \frac{a}{2} \, \Pi_{z} - \dfrac{1}{2} \, \Big[  b_{3} \, a \, x \, \sin \Big( \frac{2 z}{a} \Big) + b_{3} \, a \, y \, \cos \Big( \frac{2 z}{a} \Big) \Big] \, .
\end{aligned}
\eeq
Their commutation relations read
\begin{equation}
\{ H_{1}, H_{2} \} = 0 \, , \qquad \{ H_{1}, H_{3} \} = - H_{2} \, , \qquad \{ H_{2}, H_{3} \} = H_{1} \, .
\end{equation}
The Haantjes operators associated with this system are
 $\{\bs{K}_1, \bs{K}_2, \bs{K}_5\}$, where $\bs{K}_1$ and $\bs{K}_2$ are given by Eqs. \eqref{eq:4.7} and \eqref{eq:4.10} respectively, and
\begin{equation}
\begin{aligned}
& \boldsymbol{K}_{5} = \frac{1}{a \Lambda_z - 2 \, \left(x^2+y^2\right) \, \Pi_{z}} \\ & \qquad
\left[
\begin{array}{ccc|ccc}
 a y^2 & - a x y & 2 y \left(x^2+y^2\right) & 0 & 0 & 0 \\
 - a x y & a x^2 & - 2 x \left(x^2+y^2\right) & 0 & 0 & 0 \\
 \frac{a^2}{2} y & - \frac{a^2}{2} x & a \left(x^2+y^2\right) & 0 & 0 & 0 \\ \hline
 0 & f_{5} & g_{5} & a y^2 & - a x y & \frac{a^2}{2} y \\
 - f_{5} & 0 & h_{5} & - a x y & a x^2 & - \frac{a^2}{2} x \\
 -g_{5} & - h_{5} & 0 & 2 y \left(x^2+y^2\right) & - 2 x \left(x^2+y^2\right) & a \left(x^2+y^2\right) \\
\end{array}
\right] \, , \label{eq:op_k3_helond_nob1_cart}
\end{aligned}
\end{equation}
with
\begin{align*}
& f_{5} = \frac{a}{2} \, \left[ a \, x \, \partial_z A_x + a \, y \, \partial_z A_y + 2 \left(x \, y \, \left( \partial_x A_x - \partial_y A_y \right)+\left(y^2-x^2\right) \, \partial_x A_y - \Lambda_z \right)\right] \, , \\
& g_{5} = \frac{1}{2} \, a^2 \, y \, \partial_z A_z - a \left(x^2+y^2\right) \, \partial_z A_x + a \, y^2 \, \partial_x A_z - a \, x \, y \, \partial_y A_z - 2 \, y \, ( y^2 + x^2 ) \, \partial_x A_x \\ & \qquad \qquad + 2 \, x \, \left(x^2+y^2\right) \partial_x A_y + 2 \, \left(x^2+y^2\right) \, \Pi_{y} \, , \\
& h_{5} = -\frac{1}{2} \, a^2 \, x \, \partial_z A_z - a \left(x^2+y^2\right) \partial_z A_y + a \, x^2 \, \partial_y A_z - a \, x \, y \, \partial_x A_z + 2 \, x \, ( x^2 + y^2 ) \partial_y A_y \\ & \qquad \qquad - 2 \, y \, \left(x^2+y^2\right) \partial_x A_y - 2 \, \left(x^2+y^2\right) \, \Pi_{x} \, .
\end{align*}
Integrating the annihilators associated with the algebra $\mathscr{H}_1 = \{ \boldsymbol{I}, \boldsymbol{K_1}, \boldsymbol{K_2} \}$, we find the set of Darboux-Haantjes coordinates
\begin{equation}
\begin{aligned}
& q^{1} = x \, , \qquad p_{1} = \Pi_{x} + \dfrac{1}{2} a \, b_{3} \, \cos \Big( \dfrac{2 z}{a} \Big) \, , \\
& q^{2} = y \, , \qquad p_{2} = \Pi_{y} - \dfrac{1}{2} a \, b_{3} \, \sin \Big( \dfrac{2 z}{a} \Big) \, , \\
& q^{3} = z \, , \qquad p_{3} = \Pi_{z} \, .
\end{aligned} \label{eq:coord_diag_h1_helond_nob1_cart}
\end{equation}
In this chart, the operators of the algebra $\mathscr{H}_{1}$ read
\begin{equation}
\begin{aligned}
& \boldsymbol{K}_{1} =  \frac{1}{p_{1} - \frac{1}{2} a \, b_{3} \, \cos \big( \frac{2 q^{3}}{a} \big)} \, \text{diag} \, [1, 0, 0, 1, 0, 0] \, , \\
& \boldsymbol{K}_{2} = \frac{1}{p_{2} + \frac{1}{2} a \, b_{3} \, \sin \big( \frac{2 q^{3}}{a} \big)} \, \text{diag} \, [0, 1, 0, 0, 1, 0] \, ;
\end{aligned} \label{eq:ops_h1_helond_nob1_cyl}
\end{equation}
and the Hamiltonian function and the integrals of the motion take the form
\beq
\begin{aligned}
& H = \frac{1}{2} \Big[ \Big( p_{1} - \frac{1}{2} a \, b_{3} \, \cos \Big( \frac{2 q^{3}}{a} \Big) \Big)^{2} + \Big( p_{2} + \frac{1}{2} a \, b_{3} \, \sin \Big( \frac{2 q^{3}}{a} \Big) \Big)^{2} + p_{3}^{2} \Big] \, , \\
& H_{1} = p_{1} \, , \\
& H_{2} = p_{2} \, .
\end{aligned}
\eeq

\subsubsection{Partial separability of the helical undulator} 
We also mention that it is possible to construct a set of partial separation coordinates  \cite{CR2019SIGMA,DKNSIGMA2019,RTT2023prepr} for the helical undulator.  This set is related to an incomplete system of commuting Hamiltonians, i.e., a system possessing fewer integrals than the dimension of the configuration space. In our case, the incomplete Poisson algebra is $(H,H_3)$, to which corresponds the Haantjes algebra $\{\bs{I},\bs{K}_5\}$. The operator $\boldsymbol{K}_{5}$ is semisimple and diagonalizes in the chart
\begin{equation}
\begin{aligned}
& q^{1} = r \, , \\
& q^{2} = \frac{a}{2} \, \varphi + z \, , \\
& q^{3} = \frac{a}{2} \, \varphi - z \, ,
\end{aligned} \qquad
\begin{aligned}
& p_{1} = \Pi_{r} \, , \\
& p_{2} = \frac{2}{a} \Pi_{\varphi} + \Pi_{z} + b_{3} r \sin \Big( \varphi + \dfrac{2 z}{a} \Big) \, , \\
& p_{3} = \frac{2}{a} \Pi_{\varphi} - \Pi_{z} - b_{3} r \sin \Big( \varphi + \dfrac{2 z}{a} \Big) \, .
\end{aligned}
\end{equation}
These coordinates represent a set of partial separation variables. The integrals of the motion take the form
\beq
\begin{aligned}
& H = \frac{1}{8 (q^{1})^2} \bigg[ a^2 (p_{2}+p_{3})^2+4 b_{3} \, (q^{1})^3 \sin \Big(\frac{2 q^{2}}{a}\Big) \Big(b_{3} \, q^{1} \sin \Big(\frac{2 q^{2}}{a}\Big)-2 p_{2}+2 p_{3}\Big)+4 (q^{1})^2 \left(p_{1}^2+(p_{2}-p_{3})^2\right) \bigg] \, , \\
& H_{3} = a \, p_3 \, . 
\end{aligned}
\eeq
The corresponding system of Hamilton-Jacobi equations splits into a system of a PDE and an ODE, which a priori provide an integral $W$ for the undulator:
\begin{equation*}
\begin{cases}
a^2 \left(\frac{\partial W}{\partial q^{2}}+ h_{3} \right)^2+4 b_{3} (q^{1})^3 \sin \left(\frac{2 q^{2}}{a}\right) \left(b_{3} q^{1} \sin \left(\frac{2 q^{2}}{a}\right)-2 \frac{\partial W}{\partial q^{2}}+2 \,  h_{3} \right)+4 (q^{1})^2 \left(\left( \frac{\partial W}{\partial q^{1}} \right)^2+\left(\frac{\partial W}{\partial q^{2}}- h_{3} \right)^2\right) \\
 \qquad = 8 (q^{1})^2 h, \\ 
 \frac{\partial W}{\partial q^{3}} = h_{3},
\end{cases}
\end{equation*}
where $W(q^{1}, q^{2}, q^{3})$ is the Hamilton's characteristic function and $h$, $h_{3}$ are arbitrary constants.

\section{Relation with the Stäckel geometry: The AKN systems} \label{sec:5}
In this section, we shall establish a new connection between the Haantjes geometry for Hamiltonian systems in a magnetic field and the Stäckel geometry. In particular, we shall prove that the magnetic models discussed above belong to the generalized class of Stäckel systems introduced in \cite{AKN1997}, namely, they are AKN systems.

To prove this result, we shall determine the \textit{Sklyanin separation equations} \cite{Skl} and the associated Stäckel matrices  for two specific systems, namely the Hamiltonian superintegrable model for a constant magnetic field and the helical undulator. 


For each set of separation coordinates, there exists a corresponding set of SE and an associated St\"ackel matrix, as we shall see.




\subsection{St\"ackel analysis for the system in a constant magnetic field}
We shall distinguish three cases.
\vspace{2mm}

(I) We shall first work in the coordinates \eqref{eq:4.19} that diagonalize the Haantjes algebra $\mathscr{H}_{1}$. In order to obtain the corresponding Stäckel matrix, we need to rewrite the system \eqref{eq:4.22} with a suitable functional combination of the Hamiltonians:
\begin{align} \label{eq:5.1}
\nn & \tilde{H}_{1} = 2 \, H - H_{1}^{2} = - 2 \, b \, q^{2} \, p_{1} + b^{2} \, (q^{2})^{2} + p_{2}^{2} + p_{3}^2 \, , \\
& \tilde{H}_{2} = H_{1} = p_{1} \, , \\
\nn & \tilde{H}_{3} = H_{3}^2 = p_{3}^2 \, .
\end{align}

Now, it is easy to see that the equations
\begin{equation} 
S\left[
\begin{array}{c}
 \tilde{H}_{1} \\
 \tilde{H}_{2} \\
 \tilde{H}_{3} 
\end{array}
\right] =  
\left[
\begin{array}{c}
 f_{1} (q^{1}, p_{1}) \\
 f_{2} (q^{2}, p_{2}) \\
 f_{3} (q^{3}, p_{3}) 
\end{array}
\right] \, , \qquad
S = 
\left[
\begin{array}{ccc}
 S_{11} (q^{1}) & S_{12} (q^{1}) & S_{13} (q^{1})  \\
 S_{21} (q^{2}) & S_{22} (q^{2}) & S_{23} (q^{2}) \\
 S_{31} (q^{3}) & S_{31} (q^{3}) & S_{31} (q^{3})
\end{array}
\right] \label{eq:stackel_syst}
\end{equation}
can be solved by choosing
\begin{equation} \label{SM1}
S = \left[
\begin{array}{ccc}
 0 & 1 & 0  \\
 1 & 2 \, b \, q^{2} & -1 \\
 0 & 0 & 1
\end{array}
\right] \, ,
\end{equation}
and therefore $f_{1} (q^{1}, p_{1}) = p_{1}$, $f_{2} (q^{2}, p_{2}) = p_{2}^{2} + b^{2} \, (q^{2})^{2}$ and $f_{3} (q^{3}, p_{3}) = p_{3}^2$.
Notice that Eqs. \eqref{eq:stackel_syst} represent the SE for our Hamiltonian model \eqref{eq:5.1}.

(II) Let us consider the DH coordinates associated with the algebra $\mathscr{H}_{2}$. As before, we suitably combine the integrals of the motion to get the St\"ackel basis
\begin{align}
\nn & \tilde{H}_{1} = 2 \, H - H_{2}^{2} - H_{3}^{2} + H_{3} = p_{1}^{2} + 2 \, b \, q^{1} \, p_{2} + p_{3} + b^{2} \, (q^{1})^{2} \, , \\
& \tilde{H}_{2} = H_{2} = p_{2} \, , \\
\nn & \tilde{H}_{3} = H_{3} = p_{3} \, .
\end{align}
From  Eqs. \eqref{eq:stackel_syst}, we deduce the Stäckel matrix
\begin{equation} \label{eq:5.5}
S = \left[
\begin{array}{ccc}
 1 & - 2 \, b \, q^{1} & -1 \\
 0 & 1 & 0  \\
 0 & 0 & 1
\end{array}
\right],
\end{equation}
and the St\"ackel functions $f_{1} (q^{1}, p_{1}) = p_{1}^{2} + b^{2} \, (q^{1})^{2}$, $f_{2} (q^{2}, p_{2}) = p_{2}$ and $f_{3} (q^{3}, p_{3}) = p_{3}$.

(III) The cylindrical coordinates \eqref{eq:coord_diag_h3v3_bconst3d_cart} diagonalize the algebra $\mathscr{H}_{3}$. Adopting the same procedure as before, we obtain the St\"ackel basis
\begin{align} \label{eq:5.6}
\nn & \tilde{H}_{1} = 2 \, H - b \, H_{4} = p_{1}^{2} + \dfrac{p_{2}^{2}}{(q^{1})^{2}} + p_{3}^2 + \frac{1}{4} \, b^{2} \, (q^{1})^{2} \, , \\
& \tilde{H}_{2} = H_{4}^2 = p_{2}^2 \, , \\
\nn & \tilde{H}_{3} = H_{3}^2 = p_{3}^2 \, .
\end{align}
The St\"ackel matrix and functions are, respectively
\begin{equation} \label{eq:5.7}
S = \left[
\begin{array}{ccc}
 1 & - \frac{1}{(q^{1})^{2}} & -1 \\
 0 & 1 & 0  \\
 0 & 0 & 1
\end{array}
\right] \, ,
\end{equation}
and  $f_{1} (q^{1}, p_{1}) = p_{1}^{2} + \frac{1}{4} \, b^{2} \, (q^{1})^{2}$, $f_{2} (q^{2}, p_{2}) = p_{2}^2$, $f_{3} (q^{3}, p_{3}) = p_{3}^2$.
\subsection{St\"ackel analysis for the helical undulator}

By analogy with the previous discussion, we will work in the chart that diagonalizes the algebra $\mathscr{H}_{1}$ of the helical undulator. From the equations
\begin{align}
\nn & \tilde{H}_{1} = 2 \, H - H_{1}^{2} - H_{2}^{2} - \frac{1}{4} \, a^{2} \,  b_{3}^{2} = - a \, b_{3} \, p_{1} \cos \Big( \dfrac{2 \, q^{3}}{a} \Big) + a \,  b_{3} \, p_{2} \sin \Big( \dfrac{2 \, q^{3}}{a} \Big) + p_{3}^{2} \, , \\
& \tilde{H}_{2} = H_{1} = p_{1} \, , \\
\nn & \tilde{H}_{3} = H_{2} = p_{2} \, ,
\end{align}
using \eqref{eq:stackel_syst} we deduce the St\"ackel matrix
\begin{equation} \label{SM4}
S = \left[
\begin{array}{ccc}
 0 & 1 & 0  \\
 0 & 0 & 1 \\
 1 & a \, b_{3} \, \cos \big( \frac{2 q^{3}}{a} \big) & - a \, b_{3} \, \sin \big( \frac{2 q^{3}}{a} \big)
\end{array}
\right],
\end{equation}
and the Stäckel functions  $f_{1} (q^{1}, p_{1}) = p_{1}$, $f_{2} (q^{2}, p_{2}) = p_{2}$ and $f_{3} (q^{3}, p_{3}) = p_{3}^{2}$.

\section{New families of magnetic systems on a Riemannian space} \label{sec:6}
In this section, we show that the previous geometric construction leads to new integrable models. Precisely, we wish to solve an inverse problem: given a Stäckel matrix $S$, find the most general Hamiltonian family of systems admitting a given Haantjes web. This means that the family shares $S$ as the associated Stäckel matrix and the same $\omega \mathscr{H}$ manifold. 
Systems belonging to the same Haantjes web will automatically be separable in the same coordinate system.

The model with a constant magnetic field, being multiseparable, provides us with three different Stäckel matrices. We will generalize them in order to construct families of separable models.

(A) Let us consider the Stäckel matrix
\begin{equation} \label{GSM1}
	S = \left[
	\begin{array}{ccc}
		0 & 1 & 0 \\
		1 & b \, \lambda_1 (q^{2}) & \lambda_2 (q^2) \\
		0 & 0 & 1
	\end{array}
	\right],
\end{equation}
where $\lambda_1$ and $\lambda_2$ are arbitrary functions. It clearly generalizes the St\"ackel matrix \eqref{SM1}. We define the auxiliary integrable model
\begin{align} \label{eq:6.2}
	\nn & \hat{H}_1= - b \, \lambda_1 (q^2) \, \mu_1(q^1) \, p_1+ \mu_2(q^2) \, p_2^2 + \lambda_2 (q^2) \mu_3 (q^3) \, p_3^2 + b^2 \, \mu_4(q^2) \, , \\
	& \hat{H}_2= \mu_1(q^1) \, p_1 \, , \\
	\nn & \hat{H}_3 =\mu_3(q^3) \, p_3^2 \, ,
\end{align}
where $\mu_1$, $\mu_2$, $\mu_3$, $\mu_4$ are additional arbitrary functions. We can consider this model as a generalization of the system \eqref{eq:5.1}. The Stäckel functions for the system \eqref{eq:6.2} are $f_1= \mu_1(q^1) \, p_1$, $f_2= \mu_2(q^2) \, p_2^2+b^2 \, \mu_4(q^2)$, $f_3=\mu_3(q^3) \, p_3^2$. 
In order to construct the ``magnetic'' version of the latter model, we introduce the associated Hamiltonians 
\beq \label{eq:6.6a}
(\hat{H}_1+\hat{H}_2^2, \hat{H}_2,\hat{H}_3) \, .
\eeq
Then, using the transformation 
\begin{equation}
	\begin{aligned}
		& q^{1} = x \, , \qquad p_{1} = \Pi_{x} +\frac{b \, \lambda_1 (y)}{2 \, \mu_1(x)} \, , \quad \mu_1(x)\neq 0 \\
		& q^{2} = y \, , \qquad p_{2} = \Pi_{y} \, , \\
		& q^{3} = z \, , \qquad p_{3} = \Pi_{z} \, ,
	\end{aligned} \label{eq:6.6}
\end{equation}

system \eqref{eq:6.6a} provides us with the family of integrable magnetic models which are the goal of our construction:
\beq
\begin{aligned}
    & H= \mu_1^2(x) \, \Pi_{x}^2+ \mu_2(y) \, \Pi_{y}^2+ \lambda_2 (y)\mu_3(z) \, \Pi_{z}^2+b^2 \, \bigg(\mu_4(y) -\frac{\lambda_1^2(y)}{4}\bigg) \, , \\
	& H_1=\mu_1(x) \, \bigg( \Pi_{x} + \frac{b}{2} \frac{\lambda_1 (y)}{\mu_1(x)} \bigg) \, , \qquad H_2= \mu_3(z) \, \Pi_{z}^2 \, .
\end{aligned}
\eeq
Notice that this system admits a Riemannian metric $g^{ii}=\text{diag} \, [\mu_1^2(x), \mu_2(y), \lambda_2 (y)\mu_3(z)]$, where we assume $\mu_2(y)>0$ and the nonvanishing functions $\lambda_2 (y)$, $\mu_3 (z)$ have the same sign. The associated magnetic field is $B=\frac{b}{2} \lambda_1'(y) \sqrt{\mu_2(y)} \, \vec{e}_z$; this condition also implies that the transformation \eqref{eq:6.6} is canonical.

This system admits a semisimple Haantjes algebra of operators:
\begin{equation} 
\boldsymbol{K}_{1} = \dfrac{1}{2 \, \mu_1(x) \, \Pi_{x}} \, \bs{R}_1
\, ,
\qquad \boldsymbol{K}_{2} = \dfrac{1}{\lambda_2 (y)} \, \bs{R}_3, \qquad \lambda_2(y)\neq 0
\, ,
\end{equation}
with $\bs{R}_1$ and $\bs{R}_3$ defined as in Eqs. \eqref{eq:R1} and \eqref{eq:R3}.

\vspace{2mm}

(B) As in the previous case, the form of both St\"ackel matrices \eqref{eq:5.5} and \eqref{eq:5.7} suggests the following generalization:
\begin{equation} \label{SM3}
	S = \left[
	\begin{array}{ccc}
		1 & - \psi_1 (q^{1}) & \psi_2 (q^1) \\
		0 & 1 & 0 \\
		0 & 0 & 1
	\end{array}
	\right],
\end{equation}
where $\psi_1(q^1)$ and $\psi_2(q^1)$ are arbitrary functions. Following the same strategy (which entails some algebraic combinations of the involved Hamiltonians), we are led to the auxiliary integrable model

\begin{align} \label{eq:6.10}
	\nn & \hat{H}_1= \nu_1(q^1) \, p_1^2+ \psi_1 (q^1) \, \nu_2^2(q^2) \, p_2^2 + \psi_2(q^1) \, \nu_3 (q^3) \, p_3^2 + \nu_4(q^1) \, , \\
	& \hat{H}_2= \nu_2^2(q^2) \, p_2^2 \, , \qquad \hat{H}_3 =\nu_3(q^3) \, p_3^2 \, ,
\end{align}
with Stäckel functions $f_1=\nu_1(q^1) \, p_1^2+ \nu_4(q^1)$, $f_2=\nu_2^2(q^2) \, p_2^2$, $f_3=\nu_3(q^3) \, p_3^2$. It widely generalizes the St\"ackel basis \eqref{eq:5.6}, since $\nu_1$, $\nu_2$, $\nu_3$, $\nu_4$ are also arbitrary functions. 
To construct, as before, the ``magnetic'' version of system \eqref{eq:6.10}, we first observe that $\hat{H}_4=\nu_2(q^2) \, p_2$ is a first integral. Then, we consider the associated model
\beq \label{eq:6.12}
(\hat{H}_1+\hat{H}_4, \hat{H}_2,\hat{H}_3) \, .
\eeq
By means of the transformation
\begin{equation}
	\begin{aligned}
		& q^{1} = x \, , \qquad p_{1} = \Pi_{x} \, , \\
		& q^{2} = y \, , \qquad p_{2} = \Pi_{y}-\frac{1}{2 \, \psi_1 (x) \, \nu_2(y)} \, , \qquad \psi_1(x),\nu_2(y)\neq 0 \\
		& q^{3} = z \, , \qquad p_{3} = \Pi_{z} \, ,
	\end{aligned} \label{eq:6.13}
\end{equation}
we get the new integrable Hamiltonian family of magnetic system
\begin{align}
	& \nn H= \nu_1(x) \, \Pi_{x}^2+ \psi_1 (x) \, \nu_2^2(y) \, \Pi_{y}^2 +\psi_2(x)\nu_3 (z) \, \Pi_{z}^2 + \nu_4(x)-\frac{1}{4 \, \psi_1 (x)} \, , \\
	& H_1= \nu_2^2(y) \, \left( \Pi_{y} -\frac{1}{2 \, \psi_1 (x) \, \nu_2 (y)} \right)^2 \, , \\
	\nn & H_2 =\nu_3(z) \, \Pi_{z}^2 \, .
\end{align}
This system admits as well a Riemannian metric $g^{ii}=\text{diag} \, [\nu_1(x), \psi_1 (x) \, \nu_2^2(y), \psi_2(x)\nu_3(z)]$, when $\psi_1(x),\nu_1(x)>0$ and the nonvanishing functions $\nu_3(z)$, $\psi_2(x)$ have the same sign. The associated magnetic field is $B=- \frac{\psi_1' (x) \sqrt{\nu_1 (x)}}{2 \, \psi_1^{\frac{3}{2}} (x)} \, \vec{e}_z$; this ensures that the transformation \eqref{eq:6.13} is canonical.

Once again, we derive the associated semisimple algebra of Haantjes operators:
\begin{equation}
\boldsymbol{K}_{2} = \dfrac{2 \, \psi_1 (x) \, \nu_2 (y) \, \Pi_y - 1}{2 \, \psi_1^2 (x) \, \nu_2 (y) \, \Pi_{y}} \, \bs{R}_2
\, ,
\qquad \boldsymbol{K}_{3} = \dfrac{1}{\psi_2 (x)} \, \bs{R}_3
\, ,
\end{equation}
with $\bs{R}_2$ and $\bs{R}_3$ defined as in Eqs. \eqref{eq:R2} and \eqref{eq:R3}.
\begin{remark}
Some of the restrictions on the arbitrary functions appearing in the models previously defined can be relaxed if one wishes to define these systems on a pseudo-Riemannian geometry.
\end{remark}

\section*{Acknowledgement}

O. K. was supported by the Grant Agency of the Czech Technical University in Prague, grant No. SGS22/178/OHK4/3T/14.  The research was initiated during O. K. ERASMUS+ stay at UCM, Madrid, to which he thanks for the warm hospitality.

 D. R. N. acknowledges the financial support of EXINA S.L. The research of P. T. has been supported by the Severo Ochoa Programme for Centres of Excellence in R\&D
(CEX2019-000904-S), Ministerio de Ciencia, Innovaci\'{o}n y Universidades y Agencia Estatal de Investigaci\'on, Spain. P. T. is member of the Gruppo Nazionale di Fisica Matematica (GNFM) of INDAM.

\end{document}